# Binospec Software System

Jan Kansky[1], Igor Chilingarian[1], Daniel Fabricant[1], Anne Matthews[1], Sean Moran[1], Martin Paegert[1], J. Duane Gibson[2], Dallan Porter[2], and John Roll[1]

[1] Harvard-Smithsonian Center for Astrophysics, Cambridge, MA, USA
[2] MMT Observatory, Tucson, MA, USA
E-mail: jan.kansky@cfa.harvard.edu

**Abstract**

Binospec is a high-throughput, 370 to 1000 nm, imaging spectrograph that addresses two adjacent 8′ by 15′ fields of view. Binospec was commissioned in late 2017 at the f/5 focus of the 6.5m MMT and is now available to all MMT observers. Here we describe the Binospec software used for observation planning, instrument control, and data reduction. The software and control systems incorporate a high level of automation to minimize observer workload. Instrument configuration and observation sequencing is implemented using a database-driven approach to maximize observatory efficiency. A web-based interface allows users to define observations, monitor status, and retrieve data products.

## 1. Introduction

Binospec is a high-throughput, 370 to 1000 nm, imaging spectrograph for the f/5 focus of the MMT 6.5m telescope that addresses two adjacent 8′ by 15′ fields of view. Binospec was commissioned in late 2017 and is now available to all MMT observers. Binospec hardware design, operation and performance are described in a companion paper (Fabricant et al. 2019). Here we describe the software used for Binospec's observation planning, instrument control, and data reduction.

The Binospec software is based on software for prior instruments including Hectospec (Fabricant et al. 2005), Hectochelle (Szentgyorgyi et al. 2011), Megacam (McLeod et al. 2015), and MMIRS (McLeod et al. 2012), but includes greater automation at all levels of operation. The observation planning, scheduling, instrument control, and image processing pipeline software is designed to maximize Binospec's efficiency and scientific impact. Binospec is operated in queue mode by queue observers, to allow efficient scheduling of targets of opportunity as well as time-critical observations (e.g. exoplanet transits) while sharing weather risks. Binospec observing modes include aperture mask spectroscopy, long slit spectroscopy, and imaging.

Binospec observing begins with submission of observing proposals to either the Center for Astrophysics's or the University of Arizona's Time Allocation Committees. These committees determine how many observing nights to award for each program. Scheduling specialists at both institutions, working with the MMT Director, plan observing runs for each active instrument on a trimester basis. Successful Binospec observers are notified by email when their observations are assigned to a queue run and are given an MMT Observatory Manager (Gibson and Porter 2018) web link to a catalog page that accepts observation details. If slit mask observations are requested, the observer is directed to the BinoMask web app to design and submit these masks. The mask submission deadline is typically set to be two weeks before the queue run, with an additional week granted to submit non-mask observation requests. The day after the mask deadline, the mask designs are processed to generate files required by the laser cutter, and within a day or two the slit masks are cut and shipped to the MMT.

Prior to each night of observing, an MMT staff scientist examines the output of the MMT Observatory Manager scheduling engine, which runs daily to calculate a proposed schedule for the night. The scientist adapts the automatically generated schedule as needed, and constructs an observing plan for that night. In the afternoon, the queue observer loads the requested slit masks into the Binospec mask changer with the assistance of MMT staff. Once afternoon calibrations or observations begin, the queue observer interacts with Bobserve, Binospec's primary observer interface. Bobserve has direct access to the list of planned observations and allows the queue observer to easily construct observation and calibration sequences. If observing conditions require deviating from the planned observation sequence, the queue observer is able to access any of the approved observations to modify the plan.

Once an observation sequence is triggered, Bobserve and a number of companion GUIs allow the queue observer to send the required RA, Dec, and position angle to the telescope operator, acquire guide stars and set up guiding. The telescope operator uses a GUI to select a star for off-axis wave front sensing to correct telescope collimation and primary mirror support forces. Once guiding and wavefront sensing are underway, the queue observer initiates an exposure sequence. As a first step, the shutters are opened and the flexure control system corrects focus and removes flexure offsets. The shutters close, the CCDs are cleared, and then the desired exposures are initiated.

The raw data are almost immediately available for download on the observer's original catalog page, and are transferred to the CfA for pipeline processing. Approximately one day following the observation the processed data are uploaded and made available via the catalog page. Queue observer comments and information about the observing conditions are also available on the catalog page.

This paper describes the software required to carry out these steps except for the MMT Observatory Manager that is described in Gibson and Porter (2018).

## 2. Software Architecture

*2.1 Database Driven Observing Architecture*

    All information necessary to plan, perform, process and evaluate Binospec observations is stored in two PostgeSQL databases, one managed by the MMT staff (the Scheduler Database) and one managed by CfA staff (the Binospec Database). This architecture allows Binospec operations software to be closely coordinated with the queue scheduling operations handled by the MMT, while maintaining clear lines of responsibility between CfA and MMT software staff.  The databases are hosted on a shared server at the MMT running PostgreSQL v9.5.

    The Scheduler Database primarily supports the MMT Observatory Manager system (Gibson and Porter 2018) used observatory-wide to organize and store information about telescope schedules, observing proposals, and other information required for MMT operations. For MMIRS and Binospec the Scheduler Database contains information used to plan and execute queue observing runs, including catalogs of observation requests with object names and coordinates, requested exposure times and number of exposures, grating and central wavelength for spectroscopy, or filter for imaging.

    The Binospec Database, designed at the CfA, stores additional information needed for Binospec operations. This includes the design details for slit masks, the current state of Binospec including the installed complement of slit masks, the currently selected mask, gratings, etc.  The Binospec Database contains a timestamped list of all exposures, with the instrument configuration, queue observer notes, seeing, and the location of the data files.   Both databases are integral to Binospec observing. Figure 1 shows the observing flow and database interactions for observing with Bobserve, Binospec's observer interface.

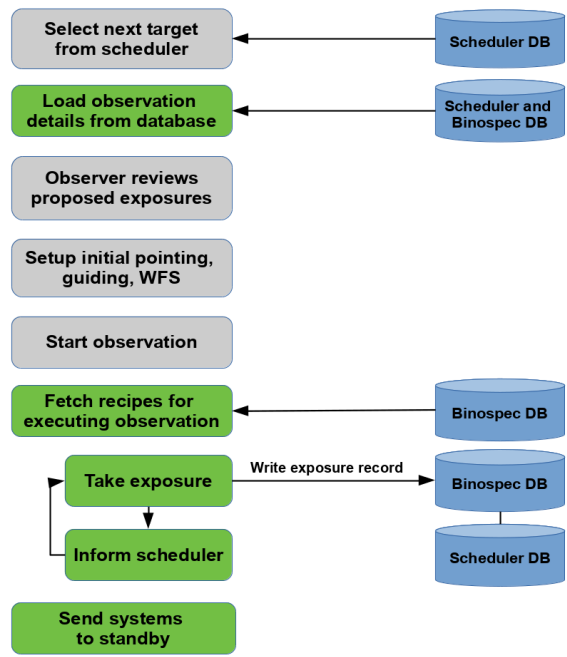

**Figure 1. Observation flow and communications with the Scheduler and Binospec Databases.  Gray blocks represent queue observer actions and green blocks represent software actions.  Binospec's observer interface, Bobserve, reads observation requests from the scheduler database and sets up a sequence of exposures for a given catalog target ID.**

    The two databases are linked through a common catalog target identification (ID) number, assigned when a user submits a queue observation request to the MMT Observatory Manager web interface. Each catalog target corresponds to one set of observations, e.g., five 300s observations of target A using mask B with instrument configuration C.  In the Scheduler Database the ID number is linked to observing programs, PI information, scheduled observing date, etc.  In the Binospec Database the ID number is linked to all images and corresponding metadata (exposure time, mask, filter, grating, FITS files, seeing, etc).

Binospec operations and the MMT Observatory Manager have read access to both databases and write access to their native databases. When a Binospec image is acquired, a record is written to the images table in the Binospec Database that includes the catalog target ID. The scheduler monitors the images table for updates, and when new exposures for a catalog target ID are found, updates the Scheduler Database with exposure count, observer comments, validity flags, FITS filenames, etc. FITS files are stored in a shared disk space visible to all software components, external to the databases. The MMT Observatory Manager web interface provides access to raw data through a link created when an exposure with that catalog target ID is recorded in the Binospec Database.

We produce a script for the data reduction pipeline and bundle it with the images (science images and all-sky camera images for cloud monitoring) for each catalog target ID at the end of the night. The bundles go to reduction servers at CfA where the data pipeline is run. The results are inspected and, if necessary, reduction parameters are adjusted and the pipeline is rerun. Once the reduced data meet our quality criteria, they are sent to the MMT server. The scheduler recognizes the new reduced images and creates a database record of the reduction for the appropriate catalog target ID and informs PIs that reduced data are available.

## 2.2 State-Based Instrument Control Architecture

An intermediate level of software servers decompose Binospec operations into defined states in a state-based component architecture. The definitions of these states, and the actions that occur during state transitions, capture the detailed sequences that are required to operate the instrument. At the lowest level of control, additional software components define interfaces and encapsulate commanding embedded controllers that control a variety of actuators and monitor sensors within the instrument. Automation of operations allows efficient and repeatable data collection. If debugging is necessary, manual Binospec control through engineering interfaces to software components at all levels is possible.

The computing and software tools adopted for Binospec build on the instrument control software for previous CfA instruments, but incorporate a new generation of techniques. The existing foundation of software tools is well suited to the internal control tasks that Binospec shares with prior instruments, and modern software tools allow us to realize an improved user interface.

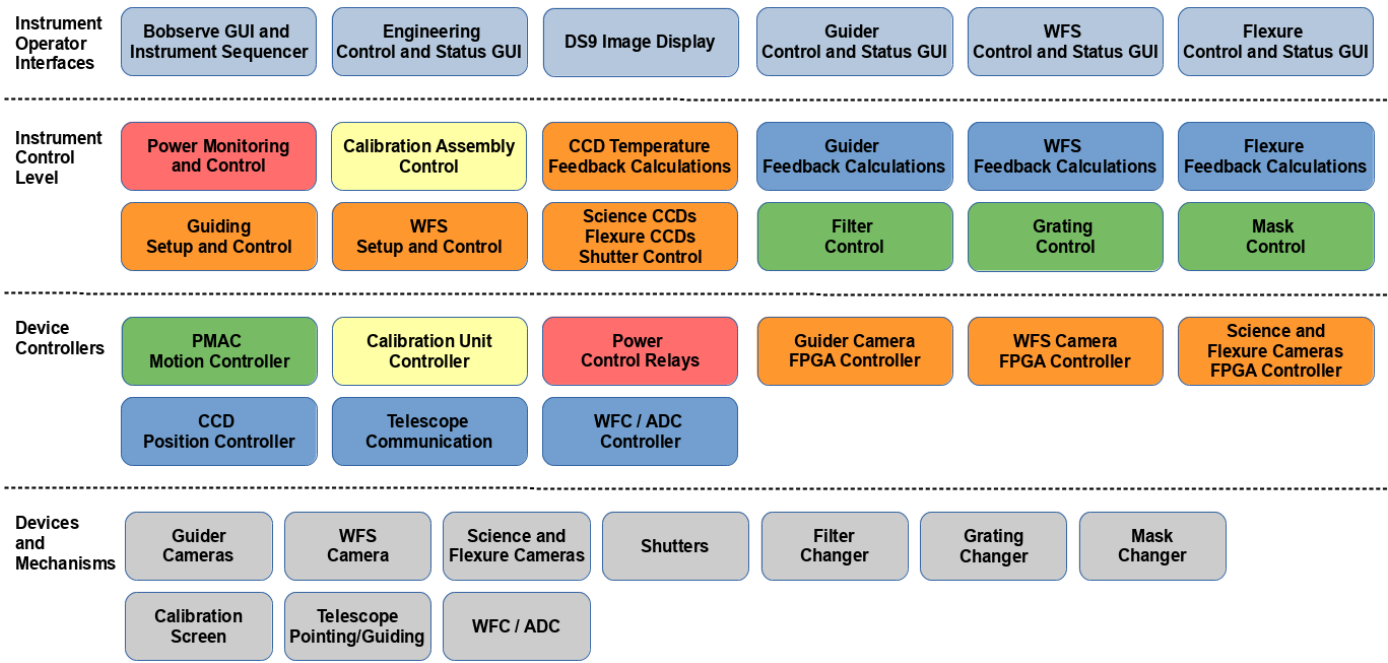

**Figure 2. Binospec control system architecture block diagram. Motion control components are green blocks, CCD camera controllers components are orange blocks, and guiding, flexure components are blue blocks, and calibration components are yellow blocks.**

The components that connect software modules through efficient inter-process communication are the core of software architecture. To maintain compatibility with existing code at the MMT, Binospec adopts the "msg" protocol as the glue between distributed components. This protocol passes ASCII messages over TCP sockets (Roll and Mandel 2001). The power of msg lies in its low-latency event-driven structure that eases management of distributed asynchronous processes. Msg provides publish-subscribe functionality, remote procedure calls, automated logging, automated connection management, and component introspection. Existing Tcl and C interfaces to the msg protocol are reused for Binospec, and a new Python interface allows access to newer software tools to implement the GUI system. The Binospec control system components are shown in Figure 2.

At the lowest levels, the Binospec control system is a collection of devices and mechanisms. At the next level, device controllers provide Ethernet interfaces to the mechanisms so that higher level processing can take place remotely. Delta Tau PMAC, Copley Accelnet, and Physik Instruminte Piezo controllers manage motion, while ADAM Ethernet I/O modules handle distributed inputs and outputs (see also the companion paper Fabricant et al. 2019).

The instrument control level coordinates the actions of the device controllers to carry out higher level mechanism control (allowing the coordination of multiple axes) and monitoring functions. The queue operator interacts with a set of displays and user interfaces that manage the functionality of the instrument control level. The settings required to define observation sequences are stored in the Binospec Database and managed by Bobserve.

*2.3 Software Tools*

The slit mask design tool, BinoMask, is a web-app built with JavaScript and HTML, using the Bootstrap V3[1] web design framework and JS9[2] image display (Mandel & Vikhlinin 2018). It makes use of JQuery[3], D3.js[4], Bootstrap-drawer[5], and a number of other packages[6] in the web front-end, as well as node.js[7] and IDL components on the server side. The Binospec operator interface is built using a Python Qt graphical interface that connects our Tcl/Tk components to a modern GUI toolkit with our existing "msg" based distributed message passing architecture (Roll and Mandel 2001). This operator interface relies heavily on PostgreSQL databases for configuration and data storage, and Redis interfaces to the MMT Observatory Manager system (Gibson & Porter 2018). Observatory Manager combines a web-based front end, a supporting REST API, a PHP web framework, an Astroplan (Morris 2016) based scheduler backend, and a Redis publish/subscribe system for inter-process communications.

The state-based layer of the Binospec software uses the SMC state machine compiler. This tool takes as input a set of Binospec state machine description files and auto-generates TclOO code that serves as the command and control framework for the middle layer of software components. Below this state-based interface, a large amount of Tcl code executes command and control actions. Tk GUIs provide engineering interfaces to many of the middle-layer components.

---

[1] http://getbootstrap.com

[2] https://js9.si.edu

[3] http://jquery.org

[4] http://d3js.org

[5] https://github.com/clineamb/bootstrap-drawer

[6] BinoMask uses suncalc.js, moment.js, d3-jetpack, bootstrap-datetimepicker, and style sheets based on sb-admin.css and font-awesome.css.

[7] https://nodejs.org

**3. Slit Mask Preparation and Observation Planning**

*3.1 Introduction*

The Binospec slit mask design tool, BinoMask, has been implemented as a web app. Any web browser can be used to design slit masks and to submit masks for manufacture with no local software installation. The user interface consists of a simple set of forms, menus, and controls framing a central JS9 (Mandel & Vikhlinin 2018) image display, which is used to visualize and manipulate the mask design projected on an astronomical image (Figure 3).

When the user specifies the RA and Dec of the mask center, BinoMask loads a Digitized Sky Survey (DSS) image of the requested region of the sky, overlaid with the Binospec field of view. The user can substitute another image that covers the correct region of the sky. The mask overlay can be clicked and dragged interactively to move or rotate the mask center and position angle. The mask design workflow follows a menu on the left-hand side of the user interface. Clicking each button expands a drawer that partially covers the image display, but reveals a series of forms and action buttons that are needed to design a mask. Starting from **Mask Configuration** at the top, a user moves through five drawers down the menu until reaching the **Submit Mask** panel at the bottom. The drawers are:

1. **Mask Configuration:** The user enters mask center coordinates, or looks up coordinates for objects in SIMBAD (Wenger et al. 2000). The user selects an initial position angle (PA), estimated date of observation, slit width and minimum slit length, and a grating/central wavelength.
2. **Target List:** The user supplies a list of potential targets from a target list in simple text format. The target list must include RA and Dec, but including object name, magnitude, and priority ranking is recommended. Targets are marked with dots on the image display.
3. **Guide Stars:** Normally candidate guide stars are selected from GAIA data release 2 (Brown et al. 2018) with a click in this panel, but a user-supplied list can be substituted.
4. **Generate Slits:** On this panel, the user enters settings for the slit layout engine, including position angle range, allowed mask center shifts, and the algorithm for optimizing slit placement (maximal slit packing, uniform wavelength coverage, or random selection, see below). Clicking Place Slits starts the optimization. When the best mask design is returned, the slit positions are displayed and slit information appears in a table.
5. **Submit Mask:** If the mask design is acceptable, this panel finishes the design and submits it to the Binospec Database. A mask design can be saved to a JavaScript Object Notation (JSON)[8] text file that can be reloaded to later complete a submission.

---

[8]https://www.json.org

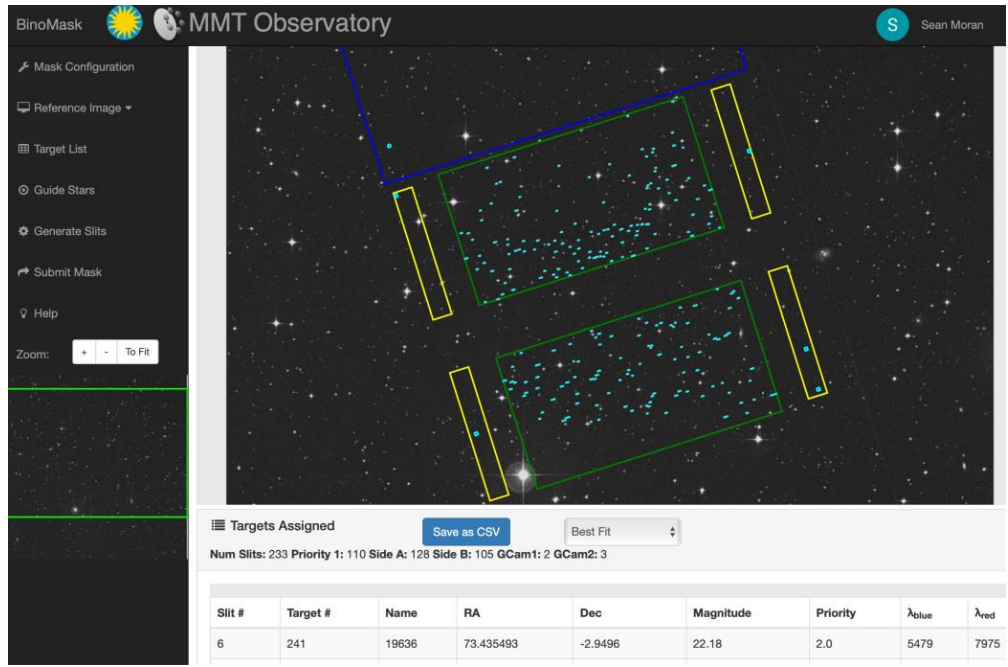

**Figure 3.** A screen shot of the BinoMask web app, with a completed mask design displayed. Clicking any item in the menu bar to the left will expand a drawer containing form fields and action buttons. The active slit mask area for science targets is outlined in green. Valid regions for guide star selection are outlined in yellow. The patrol region of the WFS camera is partly out of the image frame, but is shown in blue. Slits, guide stars boxes, and acceptable WFS stars are all outlined in cyan. Before slits have been placed, a red oval in the center(not shown) can be clicked and dragged to interactively move or rotate the mask center and PA. A table containing info on targets assigned to slits is partially visible at the bottom.

While the slit layout engine performs most calculations required to complete a mask design, we use JavaScript running in the browser to verify the observability of the mask coordinates at the requested time of year, and to estimate the best time of night to observe it. When the user selects a date within the upcoming Binospec run (prepopulated from the queue scheduler), BinoMask checks that the target is visible at airmass < 2.5 for at least one hr between 12° twilights, and calculates the best observation time.

A Binospec mask is designed for a specific date and time because differential refraction across the Binospec field of view results in small shifts (up to 0.3″) in the relative slit positions. We have developed a tool to estimate slit losses that would be incurred by observing at another time, so that queue observers can make an informed decision about changing the intended schedule.

Because the MMT rotator must rotate rapidly when a target transits near the zenith, compromising tracking, we designate masks for observation before or after transit. If the target transits during the night, the code will choose the larger of the two time windows before or after transit but the user can override this choice. If the target is rising or setting, we schedule it to begin at as low an airmass as possible.

*3.2 Slit Placement Engine*

The slit mask layout engine is an IDL routine, Fitmask_idl. Fitmask_idl constructs a grid of all allowed combinations of position and mask, and attempts to design a mask for each grid point. The position angle range is searched in steps of $1°$, and the mask (Ra, Dec) center range is searched in a five by five grid. The mask design with the highest merit function is chosen. User supplied target priorities are remapped to a 1 to 5 scale, and weighted exponentially so that the final weights range from 1 to $10^5$. The merit function for a mask is the sum of these weights for all targets placed on a mask. If multiple mask designs have equal merit functions, we choose the design closest to the nominal Ra, Dec, and position angle.

Depending on the RA, Dec and observation time, some position angles cannot be observed because the Binospec rotator limits would be reached during the observation. These invalid position angles are detected and eliminated from consideration. If no position angles within the specified range are observable, Fitmask_idl will return an error message.

Fitmask_idl places slits on a candidate mask through a series of calculations:
1. Celestial coordinates are precessed to the epoch of observation, and corrected for refraction, aberration, and nutation.
2. A sixth order polynomial is used to convert angular offsets from the field center to position offsets.
3. The slit masks are cut as a flat sheet but mounted on a mask frame that bends the mask into a cylinder along its long dimension, and tilts the mask about its long axis to best approximate the hyperbolic f/5 spectroscopic focal surface. Corrections are applied to transform position offsets in the curved focal surface to position offsets in a plane.

Next, we check whether appropriate guide stars and wavefront sensor (WFS) stars are available for the current mask center and position angle. If not, this center and position angle are rejected. If no combination of mask center and position angle in the current grid is found that provides usable guide and WFS stars, Fitmask_idl returns an error. We require a guide star with GAIA magnitude 14-18.5 in each of the two guide cameras in order to remove elevation, azimuth and rotator tracking errors. When observing, the guide stars are kept centered within 10″ square boxes. Guide stars with companions within the 10″ box that are up to three magnitudes dimmer are rejected. We require at least one WFS star (with no close companion) with GAIA magnitude 11-15 within the patrol region of the WFS camera. We normally prefer more than one guide star choice for each camera in case of error, but the user can accept a mask with a single guide star per camera.

After completing guide/WFS star selection, Fitmask_idl begins to select targets. All targets are initially assigned a slit of the requested size, and the lowest priority targets are removed until no lower priority slits overlap higher priority slits perpendicular to dispersion. Now only slits of equal priority can overlap. The user has selected among three algorithms to remove the remaining overlaps:
1. **Maximal Slit Packing:** Targets are ordered by their coordinate along the Y-axis of the mask (the spatial direction). A priority 1 target is selected at one end of the mask, and the next non-overlapping priority 1 target is selected next.
2. **Most Uniform Wavelength Coverage:** Targets are ordered by their distance from the mask midpoint in the dispersion direction (X). Target slits closer to the MMT's optical axis have a wavelength coverage shifted to the red compared to slits at the center of each Binospec beam. Slits located furthest in X from the optical axis have a wavelength coverage shifted to the blue. The slits closest to the center of each Binospec beam are retained and overlapping slit further away are removed.
3. **Randomize Selection:** This option randomizes the selection for targets of equal priority.

By default we extend the lengths of each remaining slit until overlaps would occur to gather more sky to estimate background. The user can disable this action.

*3.3 Mask Design Database*

The final step is to submit the design to the Binospec Database, and to register the design in the Scheduler Database, to allow linking to an observation request. The mask is available to the PI on their Observatory Manager catalog page, allowing the PI to generate a new catalog target, specifying the number of exposures, exposure time, and instrument setup to use when observing this mask. Mask designs saved to the database include the full state of BinoMask at time of submission, so that a mask can be displayed in a special view-only mode of BinoMask. We support editing and resubmission of an existing mask design, though that feature is not yet available to PIs.

*3.4 Mask Cutting*

As outlined in Section 2, the Observatory Manager software solicits mask designs from users in advance of each run, and sets a deadline for submission. After the deadline, a software tool called Mask2dxfqueue prepares files for the laser mask cutter. Mask2dxfqueue accesses the Scheduler Database to look up ID numbers for all masks needed for the upcoming run and accesses the Binospec Database to retrieve mask designs. Mask2dxfqueue produces DXF format files, arranging up to five mask pairs on a single 0.6 by 0.6 m laser cutter sheet. The ID number is engraved on the mask for identification when loading or unloading masks from Binospec. The masks are cut, packed, and shipped to the MMT, but we are preparing to move the laser cutter to the MMT.

## 4. Observing Interface

*4.1 Observing Queue*

The queue observer controls Binospec through Bobserve, a centralized graphical user interface (GUI) that combines three panels (Figure 4). The top panel allows the observer to monitor and manipulate software components that control Binospec subsystems. The middle panel allows the observer to specify the instrument configuration for upcoming exposures, and the bottom panel allows the observer to add exposures to the queue. A side window allows the observer to access observation details and to read PI notes from the submission.

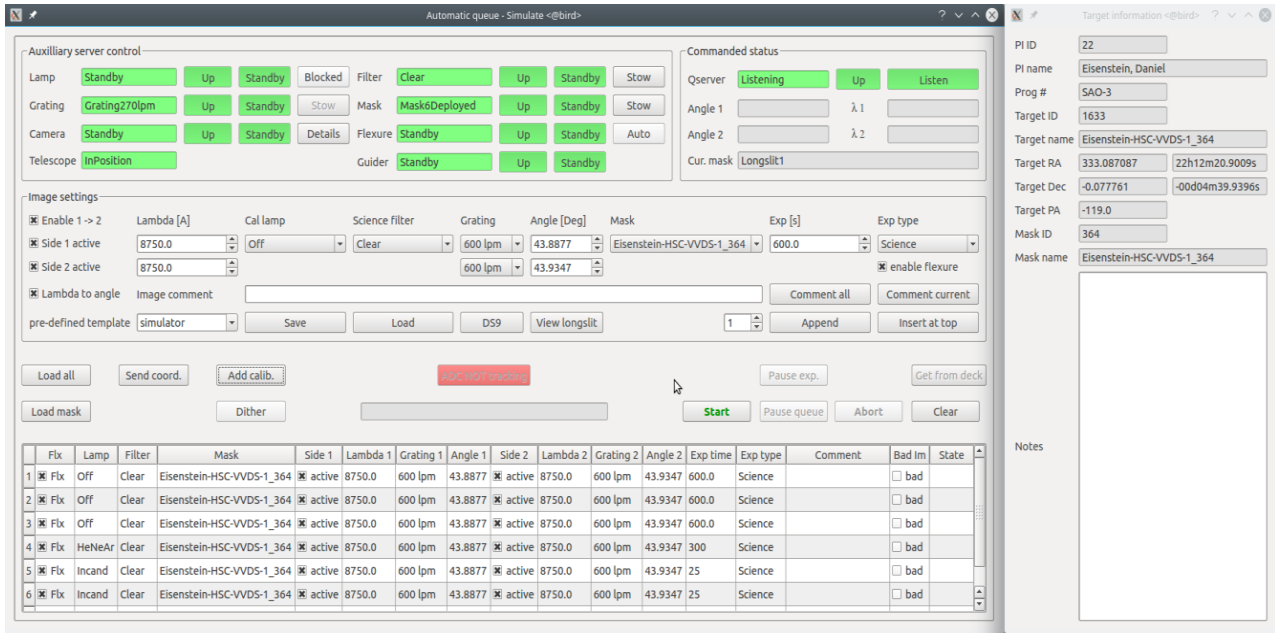

**Figure 4. Binospec queue observers interact with the instrument via a PyQt-based GUI that displays the instrument state, provides options for configuring the instrument, and displays the current actions in progress. Another window (not shown) allows the observer to record notes that will be transmitted to the PI together with the data files.**

The top instrument status panel contains indicators with red, yellow, and green colors that reflect the state of the associated subsystems. To sequence observations and calibrations, the image queue interacts with software subcomponents that encapsulate the functionality of the telescope, calibration lamps, filters, gratings, masks, guiders, the flexure control system, and the science cameras. As observation sequences proceed, the state of each subsystem is displayed within the colored indicator. This allows the operator to glance at the display and quickly establish if each subsystem is operating normally, and on more careful inspection determine the current state of the subsystem.

While the image queue can be configured manually, in standard operation the observer selects the next observing sequence from the telescope scheduler, and the unique ID for the sequence is transferred to the image queue. The queue uses this ID to query the shared database for target coordinates, number of images, desired mask, grating type, grating angles, filter type, observer, and observing program. The image queue checks the retrieved values for errors (mask not loaded, illegal central wavelength for a given grating etc.) and either displays the error condition, or proceeds to configure the table containing the sequence of exposures. Except for an exposure that has already started, the exposure time can be changed by editing the table. In addition to adjusting exposure times, one or more rows in the table can be marked and a context menu allows the following actions:
1. Inserting images before/after marked row
2. Deleting marked rows from queue
3. Pausing queue after marked row

   4. Editing comment of marked row (even after the image is completed)
   5. Viewing raw/processed image in DS9

The observer is able to adjust the preplanned observing sequence based on current observing conditions, the data quality, or any other problems with the predefined sequence.

The current exposure is highlighted when the queue is executing and a countdown in a separate window displays the remaining exposure time. An exposure in progress can be paused by temporarily closing the camera shutters. A paused exposure can be resumed or aborted as conditions dictate. Execution of the entire queue can be paused after the completion of any image in the current image queue.

On completion of an exposure, the name of the image file is added to the last column of the table, and the file is displayed in a DS9 window. If the observer inspects the image and determines that a problem exists, the entry can be flagged as bad and the database will reflect that the sequence is not yet complete. The scheduler uses the completion flags as input for its next iteration.

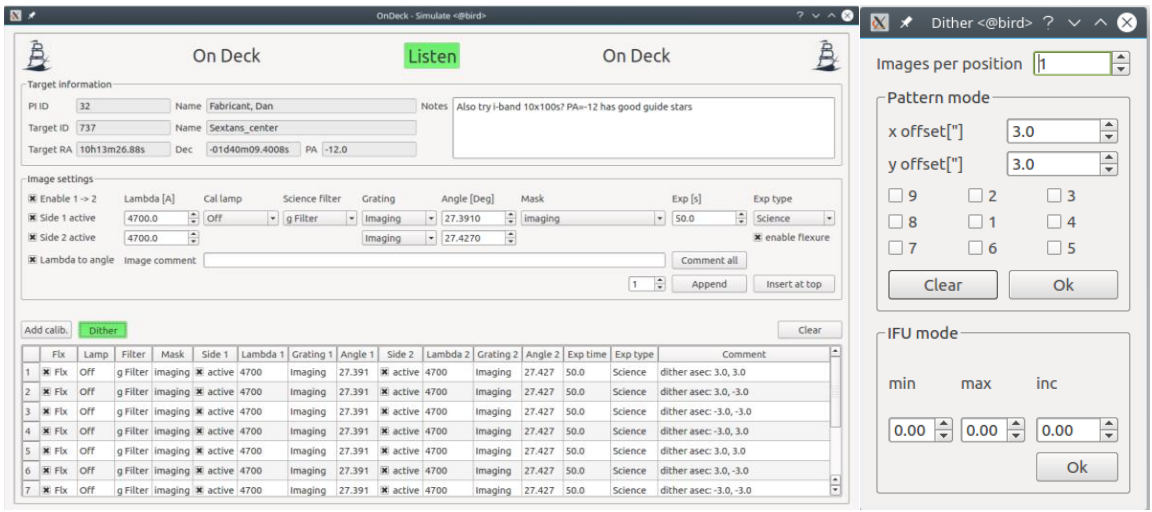

**Figure 5.** (Left) On Deck: a tool used for staging the next observation sequence while the current sequence is in progress. (Right) A dithering windows allows the observer to specify a cycle of telescope offsets that is repeated over the course of multiple observations. Additionally, a simulated IFU mode allows the observer to scan a long slit across an extended object with precise steps.

Once the queue has completed executing a complete sequence, the sequence is cleared by the observer and the system is ready to receive the next sequence directly from the scheduler or from the On Deck window (see Figure 5). The On Deck window allows the observer to validate the parameters for the next observing sequence while the current sequence is executing, rather than waiting for the current sequence to complete.

A dithering interface (Figure 5) allows the observer to specify a cycle of telescope offsets that is repeated over the course of sequential observations. Dithers in imaging mode allow better rejection of bad CCD pixels. A simulated IFU mode allows the observer to scan a long spectroscopic slit across an extended object for multiple exposures. The dithering and IFU modes are fully integrated with the guiding and wavefront systems to speed convergence after an offset, minimize operator workload, and maximize observing efficiency.

*4.2 Acquisition and Guiding*

Before the queue observer begins a sequence of observations of a new target, an acquisition sequence is activated from a guider GUI (Figure 9). Slit masks have 10″ square reference apertures to align preselected guide stars, one for each through-the-mask guider. The software identifies the boundaries of the reference apertures (illuminated by sky light) for observer approval. If approval is given, guiding corrections are sent to the telescope mount until the guide stars are centered in their boxes.

For long slit or imaging observations, individual 10″ square guide apertures are not used. Instead, each guider patrols a ~0.5′ by 16′ aperture. Guide stars from the GAIA catalog (Brown et al. 2018) are selected by the observer from a list. The guider stages are commanded to move to the correct position for the selected stars after pointing offsets to center the target on the slit (or in the imaging field of view) are issued. Subframes from the guider, centered on the correct guide star position, are read out at the selected cadence, and guiding corrections are issued to the mount until the star is centered.

As the guider is converging, the telescope operator selects a wavefront sensing star for the off-axis wavefront sensor and commands the wavefront stages to move to the appropriate position. Once the guider has converged, the wavefront CCD begins framing and the MMT's wavefront analysis system calculates collimation and primary mirror support force corrections. The observing queue is started once the wavefront corrections have converged.

**5. Instrument Control**

*5.1 Introduction*

After an exposure sequence is defined in the observing queue, the control software must efficiently execute the sequence. The software controlling the execution is split between top-level components that translate image queue sequences to a set of state machine transitions, and state machines then interact with a lower level of embedded controllers that connect directly to the hardware components.

*5.2 Top Level Control*

The instrument control procedures executed by the image queue are organized as a set of recipes. The recipes are organized in a tree-like structure that allows one recipe to be built on a collection of subrecipes. By default, all actions are executed in parallel. If necessary, recipes can be organized serially by defining a set of dependencies that govern allowed levels of concurrent execution.

Figure 6 shows RMEdit, a development tool used by the software team to construct and organize recipes. Here, the recipe for collecting a dithered image with active flexure control, flxdither, is shown. The flxdither recipe is a recipe built upon the dither and flximage recipes. The dither subrecipe sets the guider offsets, waits for guider loop convergence, and sets the position offset of the wave front sensor. The flximage subrecipe resets the flexure control system, turns on any required science image calibration lamps to maximize lamp warm-up time, deploys filters, gratings, and masks, activates the flexure compensation system, and waits for the system to achieve best focus. Once complete, any requested calibration lamps are unblocked and the science exposure is triggered. The wait list in the bottom right corner of Figure 6 specifies the relative ordering of tasks in the dither and flximage parts of the compound recipe.

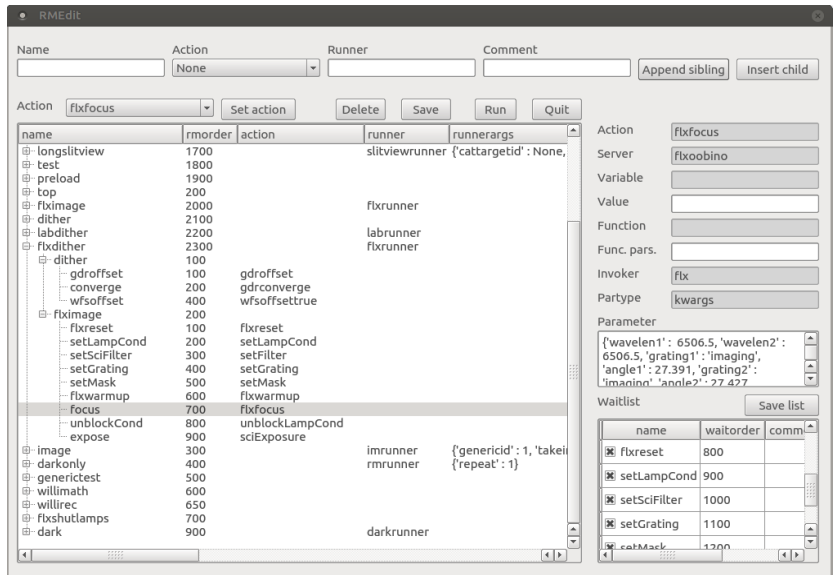

**Figure 6. RMEdit is a development tool to construct and organize control procedures (recipes). On the left an expandable tree displays existing recipes and allows creation of new recipes by a drag and drop action. Details for the marked line on the left are displayed on the right.**

Atomic actions are the most basic components of recipes that are defined using a second development tool, RMAction, that establishes the mapping between an arbitrary action name, the associated server, function call, and parameters. This tool is aware of the exposed functions and variables of all software components within the Binospec control system. These exposed functions and variables are presented to the programmer in the selection boxes at the bottom during the definition of new recipes. In this example, the gdroffset recipe is defined as a call to the gdrbino server with the name "set_tilt_offset", with the parameters "gdr_offset_x_arcsec", "gdr_offset_y_arcsec". Beyond this simple set_tilt_offset example, certain sequences of actions have special needs with regards to abort and pause signals. Aborting or pausing those sequences could leave the instrument in a state with partially deployed optical components. Those actions are executed with custom invokers that have special behavior related to the handling of pause or abort actions.

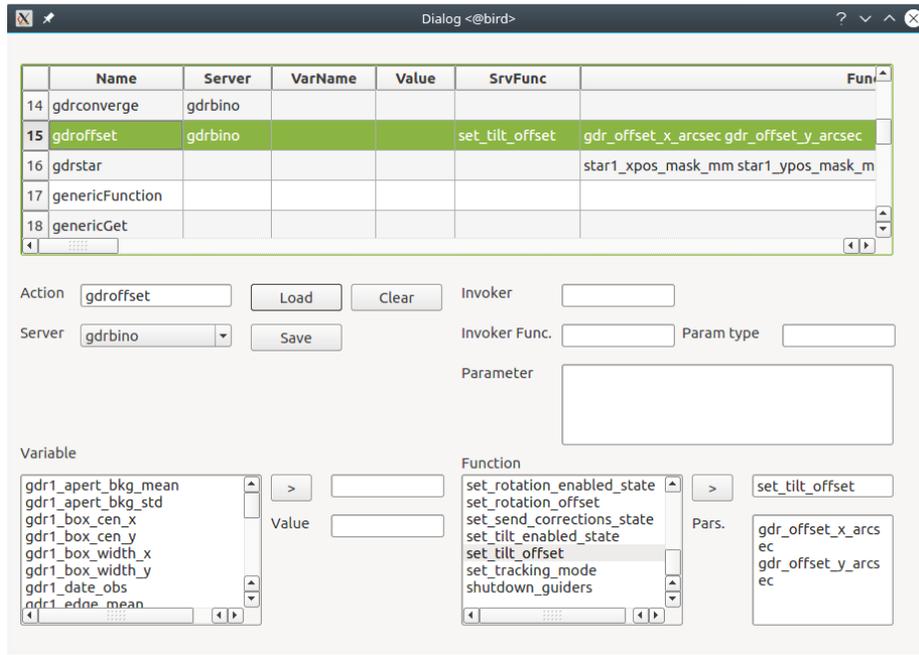

**Figure 7.** RMAction is a development tool used to develop actions, the most basic components of recipes. On top defined actions are listed. Details for the marked line are displayed at the bottom, where new actions can be defined as well.

*5.3 Low-level Control*

*5.3.1 Introduction*

At the lowest levels, the Binospec control system is a collection of power control, health and status monitoring, camera control, and motion control tasks on an Ethernet network. Commercial and custom Ethernet-based components at this level contain embedded computers that communicate with a layer of Tcl software that provides a unified msg client-server interface to these components. In cases where direct Ethernet support is not available in a controller, additional embedded computer hardware is added to provide that interface. All communication between the remote control computer and the instrument takes place over a single fiber-optic Ethernet connection.

*5.3.2 Power Control and Status Monitoring*

The power control subsystem provides filtered and overload protected AC and low voltage DC system power and communications resources to the other subassemblies of the instrument. Ethernet-controlled power strips, low voltage DC power supplies, Ethernet switchgear, solid state and electromechanical relays, terminal blocks, circuit overvoltage and current protection devices, and industrial I/O and A/D Ethernet interfaced input/output modules provide software power control and monitoring for housekeeping telemetry.

We monitor motor, electronics, and structure temperatures, as well as relative humidity, using Dallas DS18S20 digital 1-Wire temperature sensors accessed through an EDS HA7Net Ethernet to 1-Wire module. ADAM modules are used to monitor voltage and current values, along with coolant flow and air pressure. Over 160 parameters are monitored when the instrument is fully powered and the software is running.

The Binospec Operations and Monitoring Server controls system power up and power down functions, monitors system parameters, and takes action to shut the system down if an out-of-range temperature, coolant flow, or electrical parameter is detected. The monitoring software in this server checks the system parameters at a specified interval and computes a running average for each. Warning and critical ranges for the parameters are defined in ASCII files read by the server so they can be changed without rebuilding the software.

If an out-of-range value is detected, a warning or critical status is displayed on the observing GUI. For critical events, the system schedules an automatic shutdown in 15 minutes, and the user has the option to cancel the shutdown. It is also possible to disable monitoring for individual parameters if needed. A dry air control system calculates and tracks the dew point from the current temperature and relative humidity, and opens or closes a dry air valve, depending on conditions inside the instrument.

*5.3.3 Camera Control and Image Processing*

Binospec contains two science cameras with E2V 4Kx4K CCD231-84 CCDs, four 512x2048 pixel E2V CCD42-10 flexure control CCDs, two E2V 1Kx1K CCD47-20 frame-transfer guiding cameras, and a single E2V 1Kx1K frame-transfer wavefront sensing camera. All CCDs are controlled with variants of the same SAO developed camera controller based on a Xilinx Virtex-5 FX70 FPGA running embedded C code, and custom FPGA based timing for CCD device control and data acquisition. The embedded C code incorporates the Xilinx LWIP Ethernet device for providing register access to the FPGA registers for configuration and control of the CCD operation, as well as high bandwidth socket-based data delivery.

Three key pieces of Tcl software operate these cameras. The camera server, Sciflxbino, handles the task of controlling the science and flexure control CCDs. Those devices are logically grouped together from a control perspective since they must operate synchronously to prevent noise coupling from device to device in their shared dewars. The Sciflxbino server coordinates all science and flexure CCD operations to ensure that the science CCDs maintain priority, and the flexure control CCDs are read out at a fixed cadence and maintain system alignment during times when the science CCDs are exposing. Results of the science image readouts are written to disk for further processing, and also placed in shared memory for real time image quick-look with SAOImage DS9. A sample multi-object spectrum image from both science CCDs is shown in Figure 8.

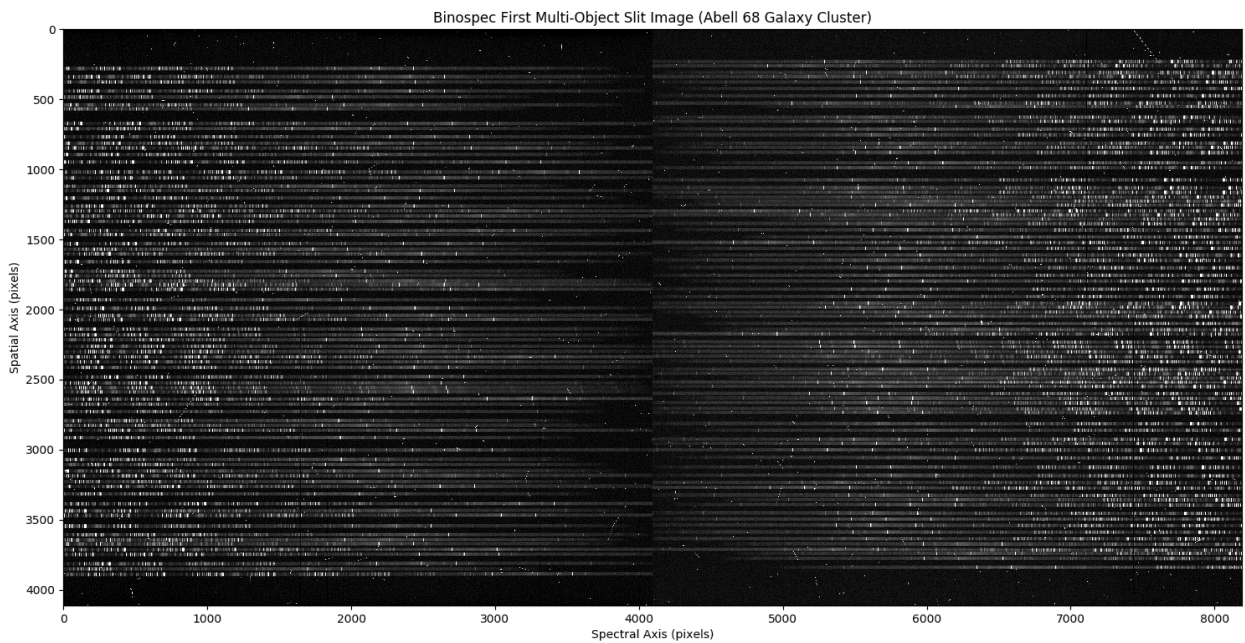

**Figure 8. A multislit spectrum for both Binospec beams. The blue end of the spectrum for both beams is at the center of the image.**

The second camera server, Gdrcambino, controls the guider camera CCDs. Three copies control the two off-axis guider cameras, and the single on-axis guider camera. When guiding, this server configures the desired subarray readout and commands the cameras into a periodic mode where the exposure timing is handled at the FPGA level. Tcl software awaits data delivery of the requested sub-array pixels over a socket connection from the FPGA, and performs pixel level calibrations of the raw images before passing the data on to higher-level guiding software.

The third camera server, Wfscambino, controls the wavefront sensor CCD. This server is similar to the guider camera control servers, but typically operates in a 2x2 binnned full-frame mode with a 30 s integration time. This server triggers new exposures with a software timer, and performs pixel level calibrations of the raw images before passing the data on to MMTO wavefront sensing software (https://github.com/MMTObservatory/mmtwfs).

Images from the off-axis guiders are received by a C-based server, Trackerbino, that handles guide star centroid processing, and mask aperture detection for through-the-mask guiding. If both guider cameras successfully detect their targets, the results of these computations are passed to a Tcl based server, Gdrbino, that converts centroid measurements in camera pixel coordinates to telescope offsets and rotator offsets that remove pointing and rotation errors. The guiding system is managed in a Qt-based GUI shown in Figure 9. This GUI allows the operator to control and monitor the guiding system. DS9 based shared-memory displays allow the operator to monitor the processed guider images with associated box and centroid positions overlays in real-time.

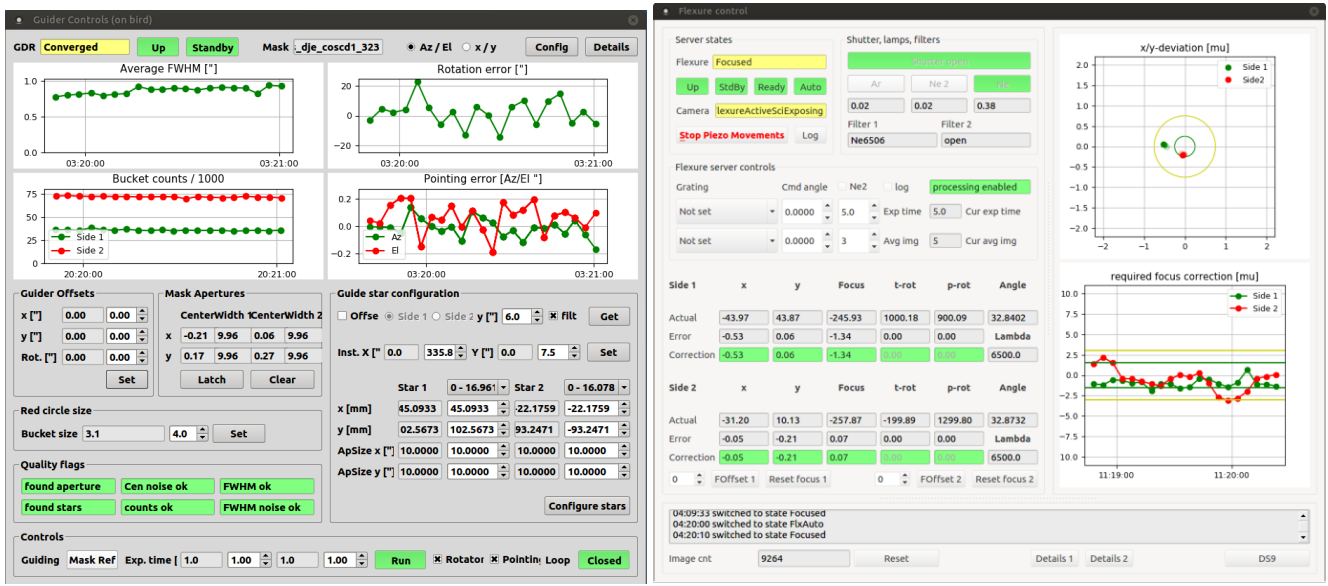

**Figure 9.** Guiding (Left) and flexure control (Right) GUIs display recent performance and provide entry boxes for configuration.

A flexure control system controls a set of calibration lamps and filters that inject light into optical fibers just behind the slit masks at the Binospec focal plane. The fibers are mounted off-axis so that they are imaged on flexure control CCDs mounted on both sides of each science CCD. Calibrations determine the desired location of these flexure control spots for each combination of grating and central wavelength. A flexure control server, Flxbino, performs the image processing and feedback to the flexure control stages to maintain this alignment. A flexure control GUI (Figure 9) allows the queue observer to monitor the performance of the flexure control system.

*5.3.4 Motion Control*

Binospec uses 21 axes of servo-controlled motors to move three guider cameras (two axes each), a wave front sensor camera (three axes), slit mask, filter and two grating selectors (two axes each), and two shutter assemblies (two axes each).  These 21 axes are controlled with a Delta Tau PMAC (Programmable Multi-Axis Controller) and Copley Controls Accelnet servo drives.

   All requests for moving these 21 axes are sent via the Tcl-based motion control server, which communicates with the PMAC over Ethernet. This Tcl server uses the Msg client interface to interpret commands and report status. The motion control server and its API are configured with an ASCII file that declares the names of the axes and maps them to the motors to which they connect. The low-level servo motion control command parameters for each axis are configured through additional ASCII files. Low-level sections of the server, for example, the moving and homing commands, are written in a Tcl-like meta-language that generates the native PMAC control language sent to the PMAC when the software is loaded.  An engineering GUI, Bbx, allows operation of the individual axes and provides detailed position information and servo status. A plotting capability is implemented in the Binospec motion control software using the data gathering capability of the PMAC. The engineering GUI can configure the data gathering before the move, command the move, and then retrieve the gathered data and plots as shown in Figure 10.  This capability is useful for initial servo tuning and subsequent performance checks.

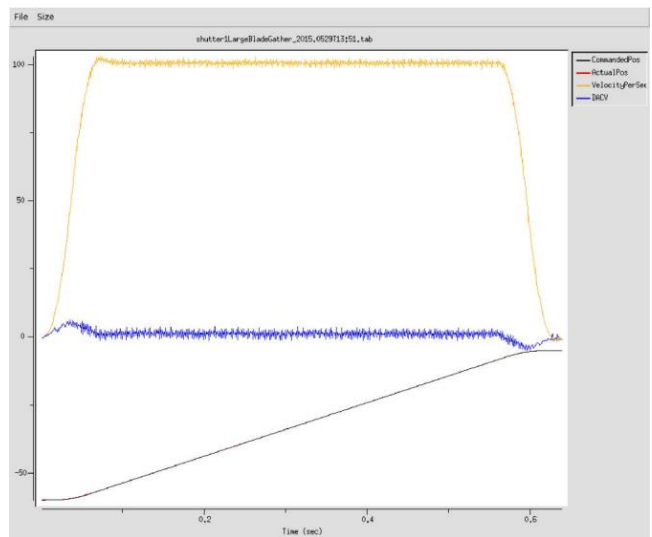

**Figure 10.  Servo performance plot for a shutter blade move of 55 mm.  The blade accelerates to a constant velocity of 100 mm s$^{-1}$ in 0.070 s and then deaccelerates in 0.070 s at the end of the move.  The velocity profile is shown in yellow, the controller output voltage in blue, and the motion profile in grey.**

*5.3.5 Changer Motion Control*

   The most complex motion control in Binospec is the operation of the slit mask, filter, and grating deployment and storage assemblies.  These device changers each include a selection device and an actuator arm that loads or stows the selected item.  The selection device is a moving elevator for masks and filters, or a rotary stage for gratings.  The grating rotary stage also sets the grating tilt angle.  The gratings are stored in fixed roundhouses.  Pneumatic pins lock the selected items in place when stored, and are unlocked to release the mask, filter, or grating when engaged with the actuator arm.  Hall effect sensors indicate whether a storage slot is occupied and additional sensors indicate when an item is fully deployed.  The software keeps track of which item is in which slot of the changer by means of configuration files that are updated when the changers are populated.

   Each group of motion axes associated with a particular changer is coordinated by a submodule of the server. These modules implement high-level commands for each assembly, allowing selection of filters, masks and gratings by name. When a filter or mask is loaded for use, the proper sequence of motions are made to return any current item to the storage unit and deploy another named item.

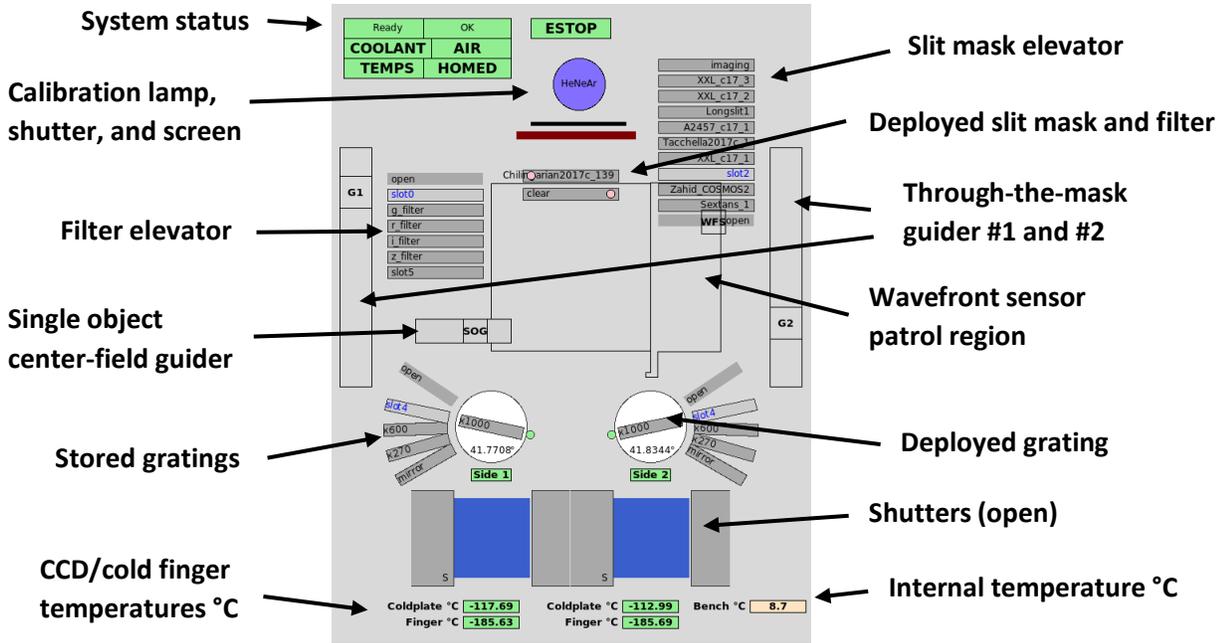

**Figure 11.** Binospec status and mechanism cartoon

Extensive safety checks are built into the motion control software to prevent collisions of these mechanisms. For example, sensors confirm that the arms and deployable items are fully inserted or retracted before allowing the storage unit to move, and are precisely positioned before the pneumatic pins are locked.

An animated Tk-based "cartoon" display (Figure 11) depicts the state of the system components, as well as the installed masks, filters, and gratings available for use. This display is connected to the motion control server for each axis and shows the relative axis positions as they move in real time, giving the queue observer visual feedback as observing configurations are changed.

*5.3.6 Flexure Motion Control*

The flexure control system commands are received and processed by a Tcl server, Pzeabino, that provides an interface between the flexure control image processing server results, and the five-axis Physik Instrumente cryogenic stages in each Binospec dewar. The companion Binospec hardware paper (Fabricant et al. 2019) gives a complete description of the flexure control system.

**6. Calibration and Data Reduction**

*6.1 Quick Spectral Focus Algorithm*

Determining the optimal focus of a wide-field spectrograph can be complex because residual axial color and off-axis aberrations must be balanced to produce the best compromise focus. Binospec's flexure control system can accurately return to a preset focus, but a large number of exposures must be analysed to calibrate the flexure control system, and we need a tool to check the focus occasionally during operation. We developed a Fourier-based spectral focus algorithm usable with long slit, multi-slit or echelle spectra.

We typically use HeAr lamp calibration frames, but the algorithm will work with red spectra with a sufficient number of sky background lines. The data image is split into a rectangular grid of subimages (8x8 by default). In each subimage the algorithm computes a fast Fourier transform (FFT) in each row along the slit and applies an optional median filter in order to remove spikes. We fit a Gaussian to the absolute value of the FFT and the result is marked "useful" if its amplitude and width exceed threshold values. All of the useful FFT arrays along the slit in the selected subimage are median combined and fit with a Gaussian. The spectral line-spread function FWHM in pixels is 2.355 L $\sigma_{comb}^{-1}$, where L is the length of the subimage in the dispersion direction and $\sigma_{comb}$ is the dispersion of the Gaussian fitted to the median FFT profile. A GUI (Figure 1Figure 12) displays the color scaled results for each subimage as well as the FWHM summary statistics (minimum, maximum, and median).

This algorithm does not require any a priori knowledge of slit positions or a wavelength calibration, and works for multi-slit and long slit spectra. The only requirements are that the slit images are approximately parallel to one of the CCD sides and that spectral lines populate all of the subimages.

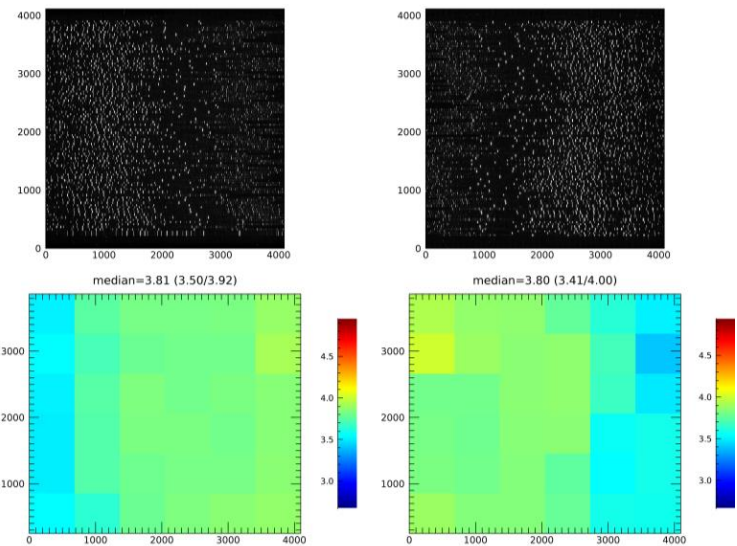

**Figure 12. The output display of the spectral focus GUI. (Top) The 270 gpm HeAr calibration images used for analysis. (Bottom) The color-scaled slit image FWHM in pixels for each region analyzed. Here the slits are 1″ wide so we expect a FWHM just under 4 pixels. The median FWHM and the FWHM range are displayed above the color plots.**

*6.2 Data Reduction Pipeline*

Binospec's IDL data reduction pipeline is based on the MMIRS pipeline (Chilingarian et al. 2015). The pipeline uses several freely distributed IDL packages: (1) the NASA Goddard ASTROLIB package (https://idlastro.gsfc.nasa.gov/), (2) the MPFIT package that implements constrained Levenberg-Marquardt minimization (Markwardt 2009) (3) mathematical procedures from SDSS idlutils (https://www.sdss.org/dr12/software/idlutils/). The pipeline supports all current Binospec configurations: three gratings at all allowed grating tilt angles with any slit mask design (multi-object or long slit) including tilted and curved slits.

The Binospec pipeline uses task control files formatted as FITS headers similarly to those used by the MMIRS pipeline. A new IDL wrapper generates the required control files from data filenames and optionally allows control over the internal algorithms.

Binospec data are stored in FITS files containing eight imaging extensions (4 read-out channels for each of two beams), and two binary table extensions that describe the mask used to acquire the data including slit dimensions, positions, and object names, etc.

We describe the data reduction steps below, emphasizing differences from the MMIRS pipeline.

*6.2.1 Primary Data Reduction*

Binospec CCDs are read out through four channels with slightly different ADU-to-electron gains, bias levels, read-out noise values, and non-linearity. We regularly measure these parameters when Binospec is in use. We subtract bias using the overscan region in every amplifier, correct for non-linearity, convert counts into electrons, and mosaic the resulting images. The two Binospec beams are then processed independently until the final spectral extraction stage. Multiple exposures are processed as a stack using an outlier resistant mean to remove cosmic ray hits. If only one frame is available, we apply the van Dokkum (2001) Laplacian cosmic ray filtering algorithm.

*6.2.2 Tracing, Extraction of 2D Slits, and Flat Fielding*

The slit mask design file allows us to predict the positions of every slit on the CCD. We match the observed flat field profile across dispersion in the center of the CCD frame in the wavelength direction against the predicted profile by varying shift and stretch. Then we repeat this procedure along the wavelength axis in order to measure geometric distortions and tilt of the spectra with respect to the CCD edge caused by the imperfect alignment of gratings in the spectrograph. Then we extract fragments of the original image containing individual slit traces and store them as individual FITS extensions for further processing. Flat fielding is done by first normalizing the flat field frame to the maximal count level in the entire frame and then dividing science images by it slit by slit. We do not remove the flat field shape from the data in order to preserve original illumination properties (e.g. vignetting).

*6.2.3 Wavelength Calibration*

We use Binospec's optical model and the 3D grating equation (Stroke 1967) to predict the positions of wavelength calibration lines for each slit, accurate to ~0.5 pixel. We measure the positions of He-Ar lines in each slit by fitting quadratic functions and then identify the lines using data from the NIST database. For multi-slit masks with more than eight slit positions we combine all line measurements on each Binospec CCD, forming a three-dimensional dataset $\lambda=F(x_{pixel},x_{mask},y_{mask})$. We fit a three-dimensional polynomial function of $3^{rd}$ order in $x_{pixel}$ and $3^{rd}$ order (270 or 600 grooves $mm^{-1}$ or gpm) or $4^{th}$ order (1000 gpm) in $x_{mask}$ and $y_{mask}$. We assign weights to individual measurements that depend on the measurement quality and inversely on the density of data points in parameter space. Weighting prevents the fit from ignoring regions with few measurements while mininizing $\chi^2$. The weighting procedure produces a smooth wavelength solution with an accuracy of ~0.025 pixel.

*6.2.4 Sky Subtraction*

The Binospec pipeline and MMIRS pipelines use similar sky subtraction algorithms. We first measure flux and position of bright airglow lines to create a two-dimensional illumination correction that we apply to each slit in the mask. For long-slit observations we use the standard Kelson (2003) sky subtraction technique; an oversampled wavelength calibrated night sky spectrum is fitted with a *b*-spline function using iterative outlier rejection and evaluated at every slit position. For multi-slit observations the sky model is generated from all slits on a given Binospec CCD, rejecting source positions if necessary. We fit *b*-splines with two additional degrees of freedom described by Legendre polynomials corresponding to X and Y slit positions. The global model is subtracted from flat-fielded two-dimensional data prior to resampling.

*6.2.5 Linearization and Flux Calibration*

After sky subtraction, we apply the wavelength solution and geometric distortion maps to generate rectified slit images, linearized in wavelength, with a preset sampling of 1.29A, 0.62A, and 0.37A for the 270gpm, 600gpm, and 1000gpm gratings, respectively. Files for each slit are stored as individual FITS extensions. The FITS headers contain information about the slit shape and size, source coordinates, and the spectral World Coordinate System. We use Binospec throughput measurements from spectrophotometric standard stars to convert reduced spectra into relative $F_\lambda$ flux units (erg $s^{-1}$ $cm^{-2}$ $nm^{-1}$). We generate preview frames by assembling the linearized spectra from each CCD into a large two-dimensional image with all of the slits aligned in wavelength. Variance frames are processed through the pipeline alongside the science frames, and are used to estimated flux uncertainties.

### *6.2.6 Extraction of One-dimensional Spectra*

We extract one-dimensional spectra from the two-dimensional linearized images. The pipeline offers two extraction algorithms: box extraction (object flux is summed along the slit with equal weights) and optimal extraction (weights reflect the distribution of flux along the slit). The latter technique yields the best signal-to-noise ratio for sources with known shape and corresponds to a matched filter in signal processing (Turin 1960). If the source profile cannot be measured accurately, for example if the spectrum has a few emission lines and no continuum, a Gaussian or Moffat profile with a specified width can be substituted. The centroid of the Gaussian profile is fit to the source data. If the centroid fit fails, the Gaussian profile is centered at the expected source position from the mask design file. If multiple sources are detected along the slit, the object closest to the mask design position is extracted.

### *6.2.7 Pipeline Code Repository*

The up-to-date pipeline code is available in the *git* repository at *bitbucket*: https://bitbucket.org/chil_sai/binospec. The pipeline is distributed under the *GPLv3* license.

## 7. Discussion and Conclusions

Increasingly complex and automated instruments require increasingly powerful software to operate efficiently and safely. We estimate that 18,000 person hours of labor were required to develop the Binospec software systems. This effort is costly, but one lost night of observing time is valued at tens of thousands of dollars, even without the lost opportunity cost to astronomers. As with any software development effort, our system strikes a compromise between reuse of existing capabilities and development of new features. While the path we have taken is specific to this development, we discuss its positive and negative aspects below to assist future efforts.

Reuse of an array of existing Tcl based code inherited from prior work at the CfA drove implementation decisions for the Binospec software. Tcl has served admirably in our past efforts, as it has for Binospec, but its age is becoming apparent and this raises concerns for future development. Most importantly, the critical mass of users required to spur new open-source projects is absent. Where gaps existed in the Tcl support libraries, we were forced to rely on our own development. As an example, cross-platform and cross-language inter-process communications and serialization libraries are commonly available in other languages, and are broadly tested and validated as parts of large collaborative open-source projects. Our reliance on our own custom-designed Msg inter-process communications library meant that extensions to support additional languages required further in-house development. This development was difficult and time-consuming. Adopting standard tools with broader support would be a better path for future efforts.

Adoption of a database-driven architecture streamlines operations and eases setting the reduction pipeline, allowing us to get data to observers more efficiently. Tight integration with our data archive is another important advantage. Although space-based observatories are usually tightly integrated with databases that track all aspects of operations, Binospec is the first MMT instrument with software designed with this approach. We cannot imagine going back to a more loosely integrated approach, and we are incorporating aspects of this database-driven operation into the software systems that support other MMT instruments. Capturing all operations in the form of database-drive recipes and sequences is a complex task requiring thoughtful design.

Structuring the Binospec control software with a state-based component architecture and a recipe-driven sequencer has improved code modularity and eased coordination between multiple developers. Describing the system by a set of meaningful states allows the queue observers to rapidly grasp system status. Earlier instrument software relied on remote procedure calls in an cross-connected web of clients and servers, leading to confusion in roles and responsibilities with poor encapsulation of low-level details. By refactoring the clients and servers as a set of hierarchical state machines, we split the roles and responsibilities into logical components, promote encapsulation of details at lower levels, and manage software complexity. We were concerned that an open-source code generation tool for the state machine layer might be an unnecessary complication, but this tool allowed us to design and modify state machines without changing the underlying code.

New web-based tools for creating observer interfaces allowed us to more easily develop software that guides observers through definition of valid observing sequences, thereby maximizing data quality and minimizing instrument configuration errors. These interfaces minimize workload for observers and reduce the support required from the Binospec team. Web-based observation planning and data access tools potentially pose support requirements because web standards and web browsers are continuously updated. Although the necessary software libraries are widely used and community supported, the level of effort required to maintain these tools over Binospec's lifetime is unknown.

The true test of a software system is in the performance, reliability, and maintainability of the deployed code. Our initial experiences in the past year indicate that our design decisions have resulted in a capable instrument that performs as expected.

**Acknowledgements**

We gratefully acknowledge the support and invaluable contributions of Stephen Amato, Nelson Caldwell, Maureen Conroy, Dylan Curley, William Joye, Eric Mandel, Brian McLeod, Matthew Smith, and Benjamin Weiner.  We thank the queue observers ShiAnne Kattner and Chun Ly for their insights and engaged feedback on Binospec operations.

**Software**

Bootstrap v3 (http://getbootstrap.com)
bootstrap-datetimepicker (https://eonasdan.github.io/bootstrap-datetimepicker)
Bootstrap-drawer (https://github.com/clineamb/bootstrap-drawer)
D3 (http://d3js.org)
d3-jetpack (https://github.com/gka/d3-jetpack)
Font Awesome (https://fontawesome.com)
jQuery (http://jquery.org)
JS9 (https://js9.si.edu)
Moment.js (https://momentjs.com)
node.js (https://nodejs.org)
PostgreSQL v9.5 (http:www.postgresql.org)
Redis (https://redis.io)
SB-Admin (https://github.com/BlackrockDigital/startbootstrap-sb-admin)
SMC state machine compiler (http://smc.sourceforge.net)
Suncalc.js (https://github.com/mourner/suncalc)